\documentclass[prd,twocolumn,nofootinbib,superscriptaddress,showkeys,citeautoscript]{revtex4}
\pdfoutput=1
\usepackage{graphicx}
\usepackage{epstopdf}
\usepackage{amsmath}
\usepackage{dcolumn}
\usepackage{multirow}
\usepackage{fancybox}
\usepackage{bm}
\usepackage{multirow}
\usepackage{array}
\usepackage{rotating}
\usepackage{ulem,fancyvrb}
\usepackage{latexsym}
\usepackage{amsfonts}
\usepackage{xcolor}
\usepackage{longtable}
\usepackage{amssymb}
\def\tana2{\tan^2\alpha_H}
\def\tanb2{\tan^2\beta}

\def\m0pr{m_0'}

\def\Mz2{M_Z^2}

\def\GeV{~{\rm GeV}}
\def\TeV{~{\rm TeV}}
\def\met100{\slashed{E}_T\geq 100~{\rm GeV}}

\newcommand{\beqn}{\begin{eqnarray}}
\newcommand{\eeqn}{\end{eqnarray}}
\newcommand{\be}{\begin{equation}}
\newcommand{\ee}{\end{equation}}
\def \cha{\tilde{\chi}^{\pm}_1}

\newcommand{\na}{\ensuremath{\tilde{\chi}^{0}_1}}

\def \nb{\tilde{\chi}^{0}_2}

\def \n34{\tilde{\chi}^{0}_{3,4}}
\newcommand{\g}{\ensuremath{\tilde{g}}}

\def \ta{\tilde{t}_1}

\def \ba{\tilde{b}_1}

\def \mhf{m_{1/2}}

\newcommand{\mh}{\ensuremath{m_{h^0}}}

\def\ra{\rightarrow}
\newcommand{\bs}{\ensuremath{B_s^0}}

\newcommand{\mm}{\ensuremath{\mu^{+}\mu^{-}}}

\newcommand{\bsmm}{\ensuremath{\bs\ra\mm}}

\newcommand{\fb}{\ensuremath{~\mathrm{fb}^{-1}}}
\newcommand{\sta}{\ensuremath{\tilde{\tau}_1}}

\def \smr{\tilde{\mu}_R}
\def \ser{\tilde{e}_R~}

\DeclareMathOperator{\sgn}{sgn}

\allowdisplaybreaks[1]

\begin{document}

\title{Higgs Boson Mass Predictions in SUGRA Unification, \\ Recent LHC-7 Results, and Dark Matter}

\author{Sujeet~Akula}  
\affiliation{Department of Physics, Northeastern University,
 Boston, MA 02115, USA}
 
\author{Baris~Altunkaynak}
\affiliation{Department of Physics, Northeastern University,
 Boston, MA 02115, USA}

\author{Daniel~Feldman}
\affiliation{Department of Physics, University of Michigan, Ann Arbor,
 Michigan, USA}

\author{Pran~Nath} 
\affiliation{Department of Physics, Northeastern University,
 Boston, MA 02115, USA}

\author{Gregory~Peim}
\affiliation{Department of Physics, Northeastern University,
 Boston, MA 02115, USA}

 \begin{abstract}
  LHC-7 has narrowed down the mass range of the light Higgs boson.  This result is consistent  with the supergravity unification framework, and the current Higgs boson mass window implies a rather significant loop correction to the tree value, pointing to  a relatively heavy scalar sparticle spectrum with universal boundary conditions.  It is shown that the largest value of the Higgs boson mass is obtained on the  Hyperbolic Branch  of radiative breaking.  The  implications of  light Higgs boson in the broader mass range of 115 GeV to 131 GeV and a narrower range of 123 GeV to 127 GeV are explored  in the context of the  discovery of supersymmetry at LHC-7 and for the observation of dark  matter in direct detection experiments. 
\end{abstract}

\keywords{ \bf Higgs, LHC-7,  Supersymmetry, Dark Matter}

\maketitle



\section{Introduction \label{intro}}
In models based on supersymmetry the light Higgs boson~\cite{HiggsBoson} has a predictive mass range,
 and recently LHC-7 has stringently constrained the light Higgs boson to lie in the
 $115\GeV$ to $131\GeV$ range~(ATLAS) and the $115\GeV$ to $127\GeV$ range~(CMS) at the $95\%$~C.L.~\cite{dec13} with possible hints of evidence within a few  GeV of $125\GeV$. 
 This mass window lies in the  range predicted by supergravity unification  {(SUGRA)}~\cite{can}
 (for reviews
 see~\cite{Nath,IbanezRoss,BSM}). In this work we 
 investigate supergravity model points that are consistent with the mass range given by  the new  LHC-7 data~\cite{dec13} (for a previous work on the analysis of the Higgs boson in SUGRA and string models
pointing to a heavier Higgs in the $120\GeV$ range
see~\cite{fln-higgs}). 

LHC-7 has made great strides in exploring the parameter space of supersymmetric models.
Indeed,  early theoretical projections   for the  expected reach in sparticle masses and 
in the  $m_0-m_{1/2}$   plane for LHC-7~\cite{arXiv:0912.4217,arXiv:1002.2430,Baer:2010tk,arXiv:1008.3423}
have been met and exceeded by the $1\fb$ and $2\fb$  LHC-7 data~\cite{cmsREACH,AtlasSUSY,atlas0lep,atlas165pb,atlas1fb}.
{ The implications of the new LHC results 
 have been analyzed}  by a number of authors in the context of lower limits on  supersymmetric
 particles and in connection with dark matter~\cite{LHC-7,Akula:2011zq,Akula:2011dd,Akula:2011ke,Akula2,Grellscheid:2011ij,Ellwanger:2011mu}. 
  Now the most recent results from CERN~\cite{dec13} indicate that
 { the two detectors,  ATLAS and CMS, have collected  as much as  $5\fb$ of data.}
  One of the most interesting implications of the LHC-7 data concerns the constraints it 
imposes on the Higgs boson  mass.

 As mentioned above we
 will work within the framework of a supergravity grand unification model
 with universal boundary conditions~\cite{can,hlw,nac}.
  Here we discuss the dependence of the  light Higgs boson mass on the parameter space,
  i.e., on $m_0, m_{1/2}, A_0, \tan\beta$~\cite{an1992},  
  where $m_0, m_{1/2}$ and $A_0$ are  the parameters at  
the GUT scale,  
where the GUT scale, { $M_\mathrm{GUT} \sim 2 \times10^{16}\GeV$} is defined as the scale at which the gauge couplings unify, and where $m_0$ is  
 soft scalar mass, $m_{1/2}$,  the gaugino mass, $A_0$, the trilinear coupling and
$\tan\beta$,  the ratio of the two Higgs VEVs in the minimal supersymmetric standard model. 

  An important aspect of SUGRA models is that the radiative electroweak symmetry breaking, REWSB, 
 is  satisfied for  $A_0/m_0$ {typically in the $-5$ to $5$ range}. The renormalization group evolution then leads
 to a value of the trilinear coupling, $A_t$, at the electroweak scale to  also be  $\mathcal{O}({\rm TeV})$.
 The relevance of this observation is that quite generically 
 supergravity unification { leads }  to a sizable  
 $A_t$    which is needed to give a substantial leading order  
  loop correction to the Higgs Boson mass  for any fixed $\mu, \tan \beta$ and $m_0$, where $\mu$ is the Higgs mixing parameter in the superpotential.
  Thus  a generic prediction of SUGRA  models under radiative electroweak symmetry breaking  for  a sizable 
  $A_0/m_0$  is that there would be a substantial loop correction to the Higgs boson mass, and it is well known that the 
light Higgs mass at the tree level has the value   $\mh \leq M_Z$ and there is a 
significant loop correction  $\Delta \mh$ to lift it above 
$M_Z$~\cite{1loop,hep-ph/9210242,Casas:1994us,Carena:1995wu,Espinosa:2000df,Carena:1998wq,hep-ph/0503173}.

The dominant one loop contribution arises from the top/stop sector and is given by
\beqn
\Delta \mh^2\simeq  \frac{3m_t^4}{2\pi^2 v^2} \ln \frac{M_{\rm S}^2}{m_t^2} 
+ \frac{3 m_t^4}{2 \pi^2 v^2}  \left(\frac{X_t^2}{M_{\rm S}^2} - \frac{X_t^4}{12 M_{\rm S}^4}\right)~,
\label{tloop}
\eeqn
where  $v=246\GeV$, $M_{\rm S}$ is an average stop mass, and $X_t$ is given  by
  \beqn
 X_t\equiv A_t - \mu \cot\beta~.  
 \label{xt}
   \eeqn
   From Eq.~(\ref{tloop}) one finds that the loop correction is maximized 
 when   
   \be X_t \sim \sqrt 6 M_{\rm S} ~.
  \label{xt}
  \ee 
We note that 
 there can be  important  loop corrections also from the 
  $b$-quark sector and a correction similar to Eq.~(\ref{tloop}) can be written
 where $X_t$ is replaced by $X_b= A_b- \mu \tan\beta$ along with other appropriate replacements. Thus when 
   $\mu \tan\beta$  becomes large,  the $b$-quark contribution
to the loop correction, which is proportional to powers of $X_b$, 
 becomes large and is comparable to the top contribution which
  implies  that a high Higgs mass can also result in
stau-coannihilation  models where typically $\mhf$ is large  and  $m_0$ is relatively small. 

 Further, we note that the  approximation  of Eq.~(\ref{xt}) would not hold if the off-diagonal elements of the stop mass squared matrix are comparable to the diagonal elements  which can happen for very large $A_t$. 
  In addition, it is well known that the two loop corrections are substantial (see e.g.~\cite{Slavich2} for a numerical analysis). While the correction at the one loop
    level has the symmetry $X_t \to - X_t$, this symmetry is lost when the two loop corrections are included 
    and then $\sgn\left(A_0/m_0\right)$ plays an important role in the corrections  to the Higgs boson mass.
    As seen later
   this observation is supported by the full numerical analysis which includes the two loop 
  corrections. We note in passing that the theoretical predictions for the light Higgs boson mass
  depend sensitively on  the input parameters  which include the gauge coupling constants as well
  as the top mass with their experimental errors. Additionally, there are also inherent
  theoretical uncertainties which together with the uncertainties of the input parameters allow
  theoretical predictions of the light Higgs boson mass  accurate to only within an error corridor
  of a few GeV (see e.g.~\cite{Slavich2}).

        Since the loop corrections involve the sparticle spectrum, 
 a large loop correction implies a relatively
heavy sparticle spectrum and specifically heavy scalars. Such a possibility 
arises in 
REWSB which allows for scalars  heavier than $10\TeV$~\cite{Chan:1997bi}.
Specifically, with scalars approaching $10\TeV$,  the Higgs boson mass
can remain heavy while the gaugino sector is free to vary. This occurs 
 within the minimal SUGRA framework and similar situations arise in 
 other works of radiative breaking~\cite{Feldman:2011ud,Baer}.

Indeed, quite generally in SUGRA and string models with the MSSM field content,  the analysis of the Higgs mass
with loop {corrections}  under the constraints 
of REWSB
 { gives an upper}
limit on the light Higgs  boson mass of about  { $135\GeV$ for a  wide range of input parameters}.\footnote{  
We note that  heavier Higgs boson  masses 
 can be obtained in  a variety of different models such as
hierarchical breaking models~\cite{Wells1,adgr,Kors:2004hz} (for recent work see~\cite{Cabrera:2011bi,arXiv:1108.6077})
or by addition of vector like multiplets~\cite{Babu:2008ge}.}
 {A very} interesting aspect of the  recent LHC-7 data concerns the fact that a large 
portion of the Higgs boson mass window has been excluded and what remains is 
consistent with the  range predicted by the SUGRA models.


  \section{ Higgs Mass in minimal SUGRA}
We discuss now the dependence of the  light Higgs boson mass on the SUGRA parameter space.
The numerical analysis was done using  a uniformly distributed random scan  over the soft parameters with $\sgn\left(\mu\right)=1$, $m_{1/2} < 5\TeV$, $\left|A_0/m_0\right| \leq -8$,  $\tan\beta \in(1,60)$ and two different ranges for $m_0$. One scan was done sampling over {lower} values of $m_0$, i.e.  $m_0\leq 4\TeV$, and has roughly $10$~million mSUGRA model points (where a model point is defined
as 1 set of the mSUGRA input parameters).  The other scan was done sampling over {larger} values of $m_0$, i.e. $m_0\geq 4\TeV$, and contains approximately $24$~million mSUGRA {model points}.  For the scan sampling over large values of $m_0$ we have imposed the upper bound of $m_0=100\TeV$.

Experimental constraints were then applied to these mSUGRA {model} points which include the limits on sparticle masses from LEP~\cite{pdgrev}:
$m_{\sta} > 81.9\GeV$, 
$m_{\cha} > 103.5\GeV$, 
$m_{\ta} > 95.7\GeV$, 
$m_{\ba} > 89\GeV$, 
$m_{\ser} > 107\GeV$, 
$m_{\smr} > 94\GeV$, 
and $m_{\g} > 308\GeV$.
Additionally, we apply the WMAP~\cite{WMAP} $4\sigma$ upper bound, i.e. $\Omega_\chi h^2 < 0.1344$.  
We define $(\Omega_\chi h^2)_\mathrm{WMAP} \equiv 0.1120$, the central value from the WMAP-7 data.  Only taking the WMAP upper limit
allows for the possibility of multicomponent dark matter~\cite{arXiv:1004.0649}.
Other constraints applied to the mSUGRA parameter points include the $g_\mu -2$~\cite{Djouadi:2006be} constraint $\left(-11.4\times 10^{-10}\right)\leq \delta \left(g_{\mu}-2\right) \leq \left(9.4\times10^{-9}\right)$ and  constraints {from} 
B-physics~{measurements}~\cite{bphys,cmslhcbbsmumu,Abazov:2010fs} {which yield flavor constraints from the data}, i.e. 
$\left(2.77\times 10^{-4} \right)\leq {\mathcal{B}r}\left(b\to s\gamma\right) \leq \left( 4.37\times 10^{-4}\right)$ 
(where this branching ratio has the NNLO correction~\cite{Misiak:2006zs}) 
and
${\mathcal{B}r}\left(B_{s}\to \mu^{+}\mu^-\right)\leq 1.1\times10^{-8}$.  As done in~\cite{Akula:2011ke,Akula:2011jx}, we will refer to these constraints as the  {\it general constraints}.  These constraints were
 imposed using {\sc micrOMEGAs}~\cite{belanger} for the  relic density as well as  for the indirect constraints 
and {\sc SoftSUSY}~\cite{Allanach}   
for the sparticle mass {spectrum}.
The {model} points are generated with {\sc SoftSUSY} version 3.2.4 which includes an important bug fix for heavy scalars when computing $\mh$.

\begin{figure}[t!]
\begin{center}
\includegraphics[scale=0.075]{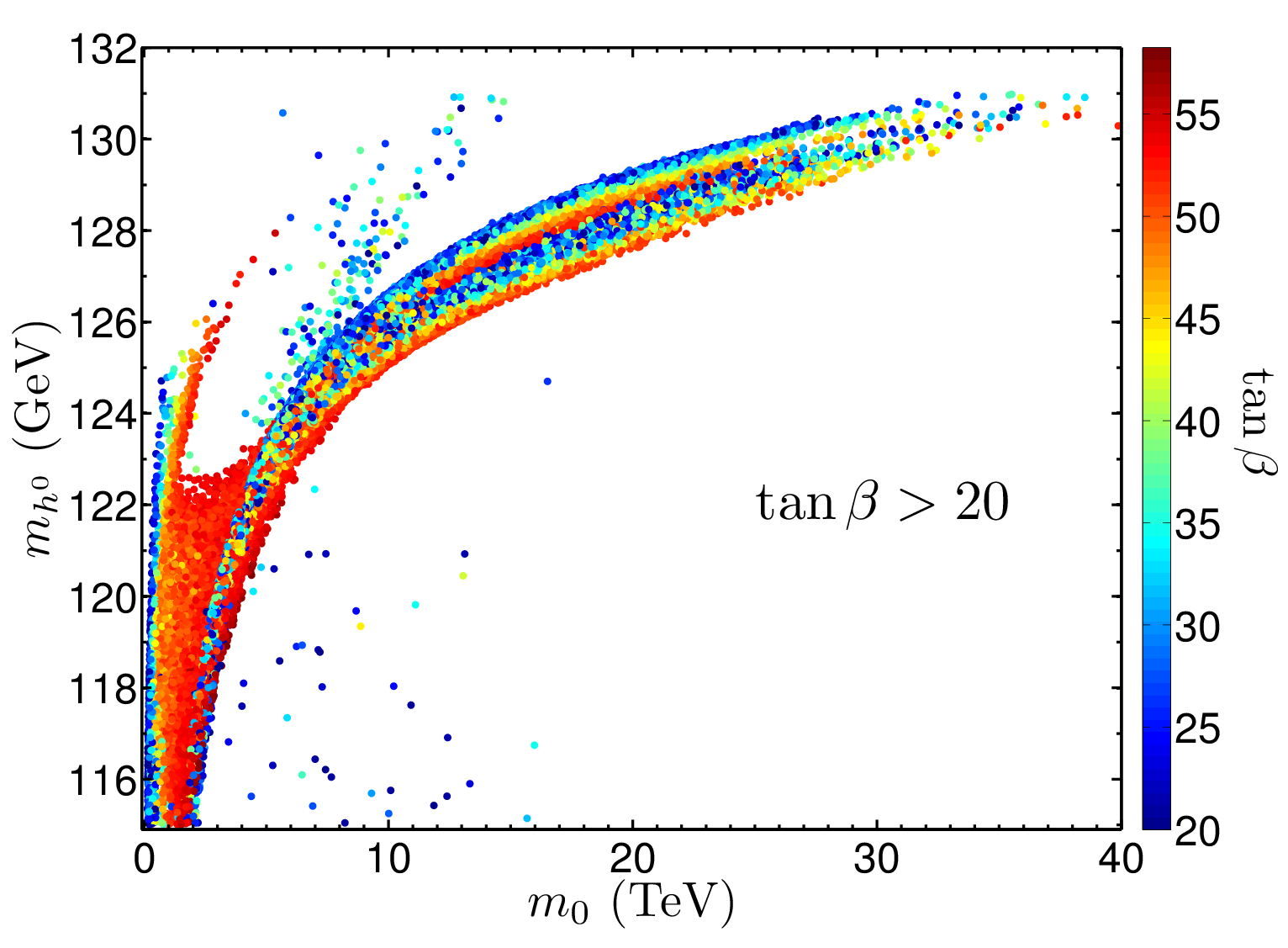}
\includegraphics[scale=0.075]{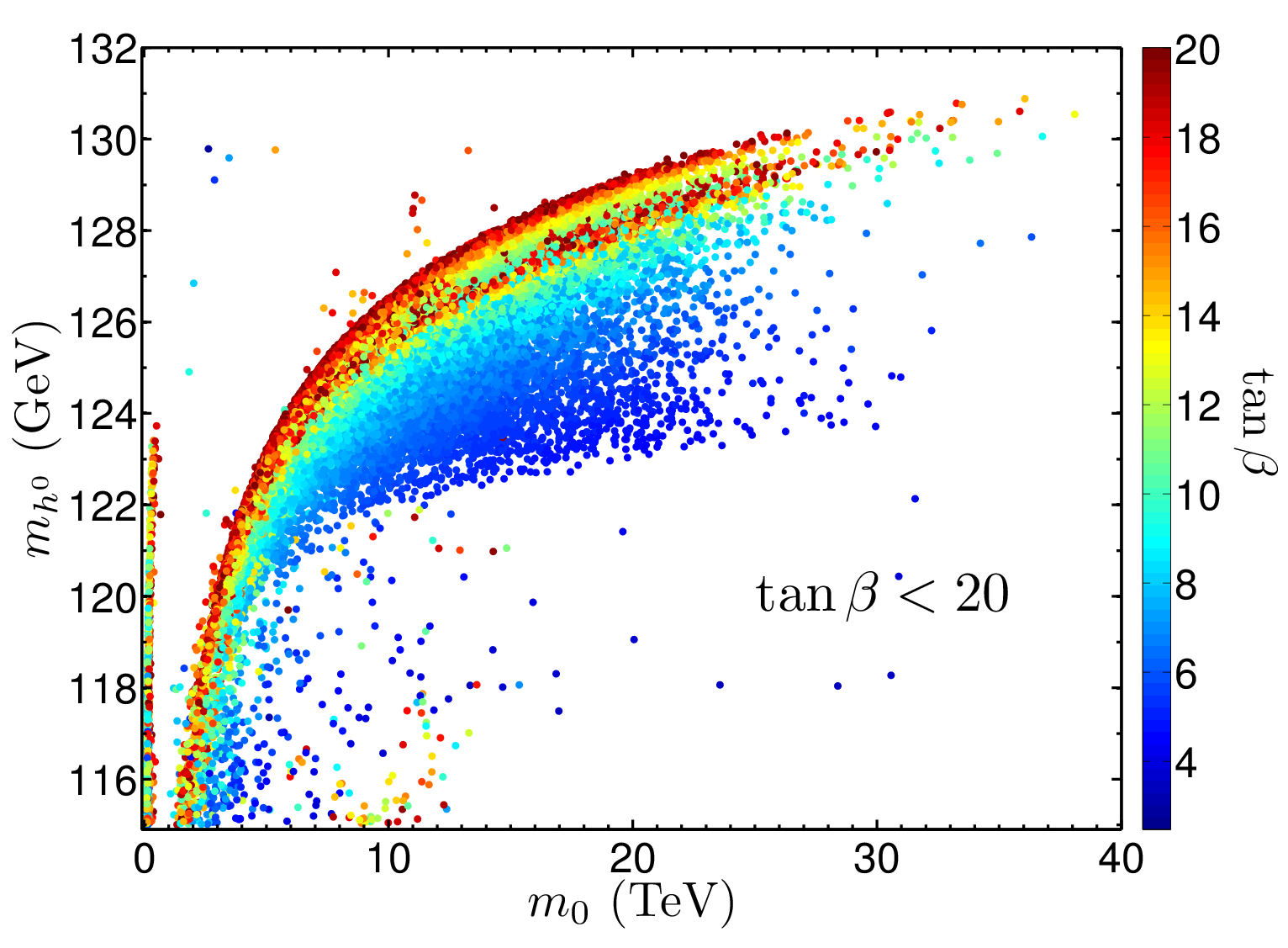}
\caption{
(color online) 
Left: 
Exhibition of the light Higgs mass as a function of $m_0$ for $\tan\beta>20$. { Right:} Same as the left panel except that 
 $\tan\beta < 20$.   
The data analyzed passes the {\it general constraints} and are generated with both scans of $m_0$. 
 \label{higgs}
 }
\end{center}
\end{figure}

\begin{figure*}[t!]
\begin{center}
\includegraphics[scale= 0.099]{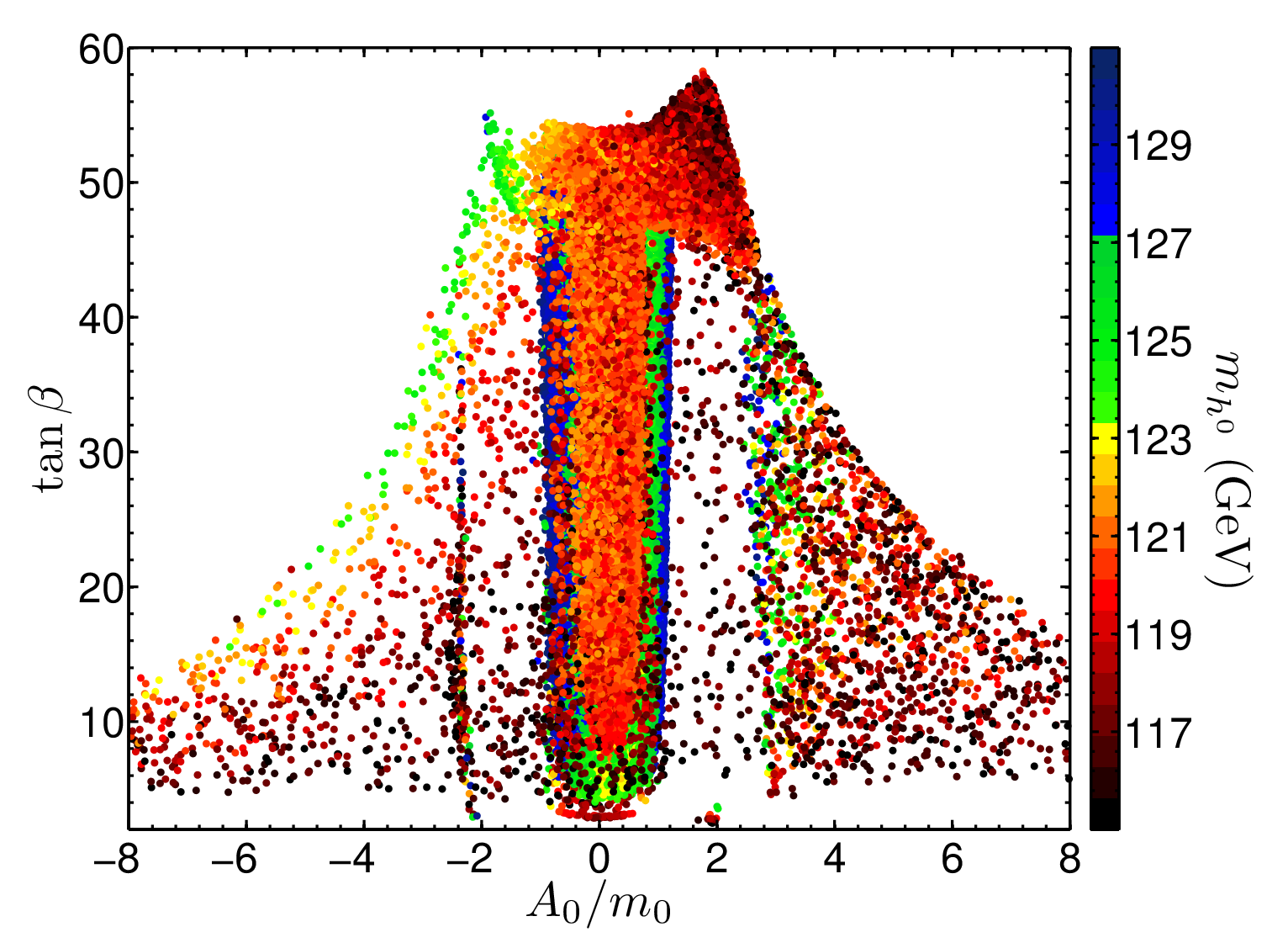} \hspace{.2cm}
\includegraphics[scale= 0.099]{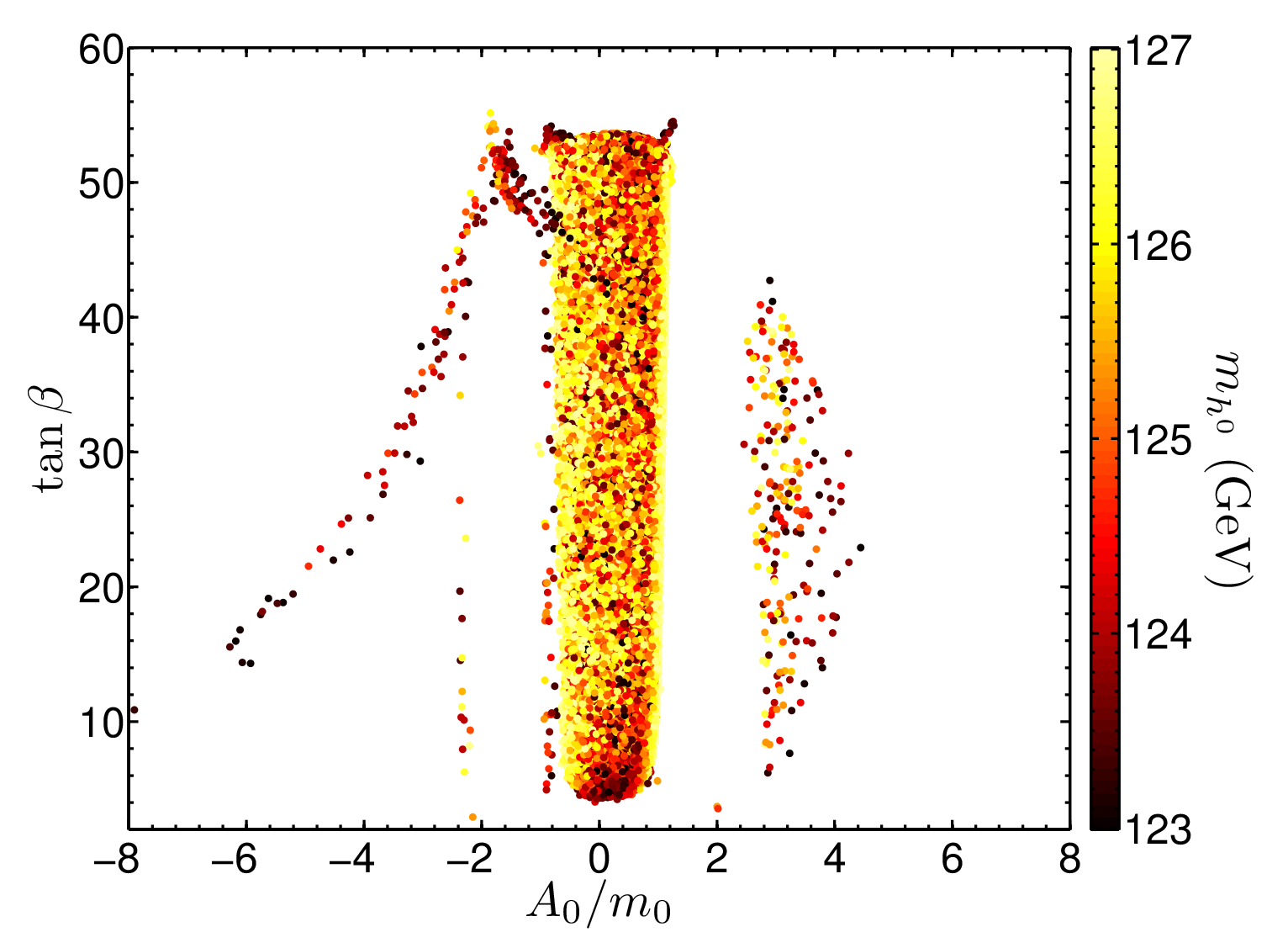}\hspace{.2cm}
\includegraphics[scale= 0.099]{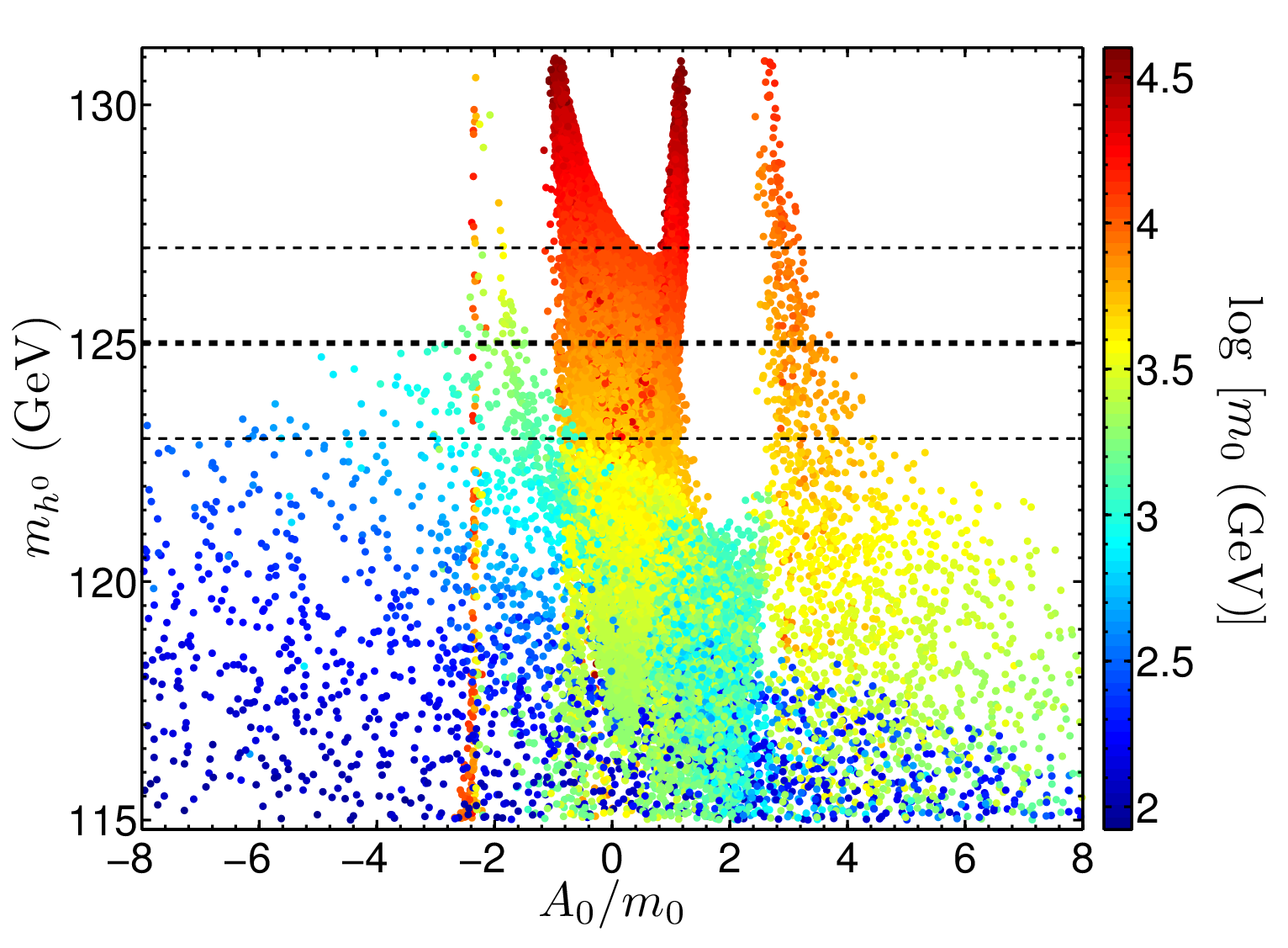}
\caption{
(color online) 
Left: 
A display of the model points  in the $\tan\beta-A_0/m_0$ plane when $\mh>115\GeV$.
Model points are shaded according to their light Higgs  boson mass, $\mh$.  
Middle: Same as the left panel except that $\mh>123\GeV$ . 
Right: Exhibition of the model points in the $\mh-A_0/m_0$ plane displayed by $\log\left(m_0\right)$
with $m_0$ in GeV units.  It is seen that 
 for low values of $|A_0/m_0|$  { larger $m_0$  corresponds to a } heavier light Higgs boson.  
The data analyzed passes the {\it general constraints} and are generated with both scans of $m_0$ {as discussed in the text}. 
 \label{higgs2}
 }
\end{center}
\end{figure*}

We display the {model} points consistent with the {\it general constraints} in Fig.~\ref{higgs} and in Fig.~\ref{higgs2}. 
In the left panel of Fig.~\ref{higgs} we exhibit the Higgs boson mass as a function of $m_0$ for 
the case when $\tan\beta > 20$ and in the right panel we exhibit it for the case when $\tan\beta < 20$. 
In both cases  we see a slow logarithmic rise  of $\mh$  with $m_0$ for large $m_0$. 
 In the left and middle panels of Fig.~\ref{higgs2} we show the distribution of the light Higgs boson mass in the $\tan\beta -A_0/m_0$ plane.  One finds that a large part of the parameter space
exists where the Higgs boson mass lies in the range $\mh>115\GeV$ (left panel) or  in the narrower range $\mh>123\GeV$ (middle panel).  In the right panel of Fig.~\ref{higgs2}, we show the distribution of $\log(m_0)$
(where $m_0$ is in GeV units)
 in the $\mh-A_0/m_0$ plane.  

{Our analysis shows a range of possibilities where a heavier Higgs boson,  i.e. $\mh\gtrsim 125\GeV$, 
can arise in the minimal supergravity model.  Thus for values of $m_0 < 4 \TeV$ a heavier Higgs boson mass can be 
gotten for a  large $A_0/m_0$ (typically  of size  $\pm 2$ with a significant  
spread). For values of $m_0>4 \TeV$  a heavier Higgs boson mass  for 
relatively smaller  values of $A_0/m_0$ is also allowed. For this case the first and second generation
sfermions may be difficult to observe while the third generation sfermions would still be accessible.
However, for the first case where a Higgs mass $\mh\gtrsim 125\GeV$  arises for  low $m_0$  
and relatively  larger $|A_0/m_0|$, the observation of signals arising from  the production of first and second generation sfermions and heavier SUSY Higgses remain
 very much within reach of the LHC with sparticles of relatively low mass in the spectrum, and variable mass hierarchies present  \cite{landscape} .  This will be shown in more detail in the next section.
}

\section{Sparticle Spectra and Higgs Mass}

 \begin{figure*}[t!]
\begin{center}
\includegraphics[scale=0.1]{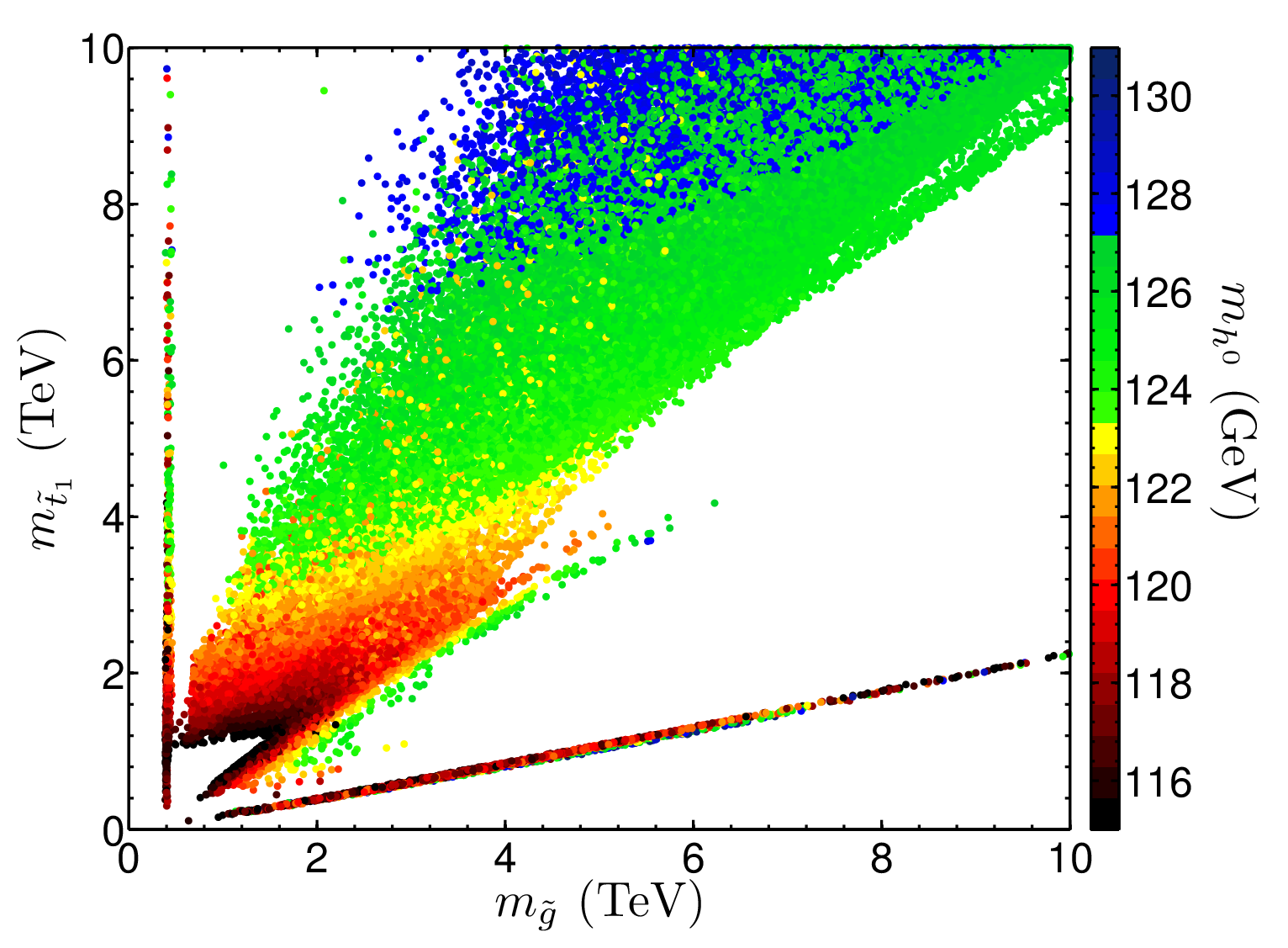}\hspace{.2cm}
\includegraphics[scale=0.1]{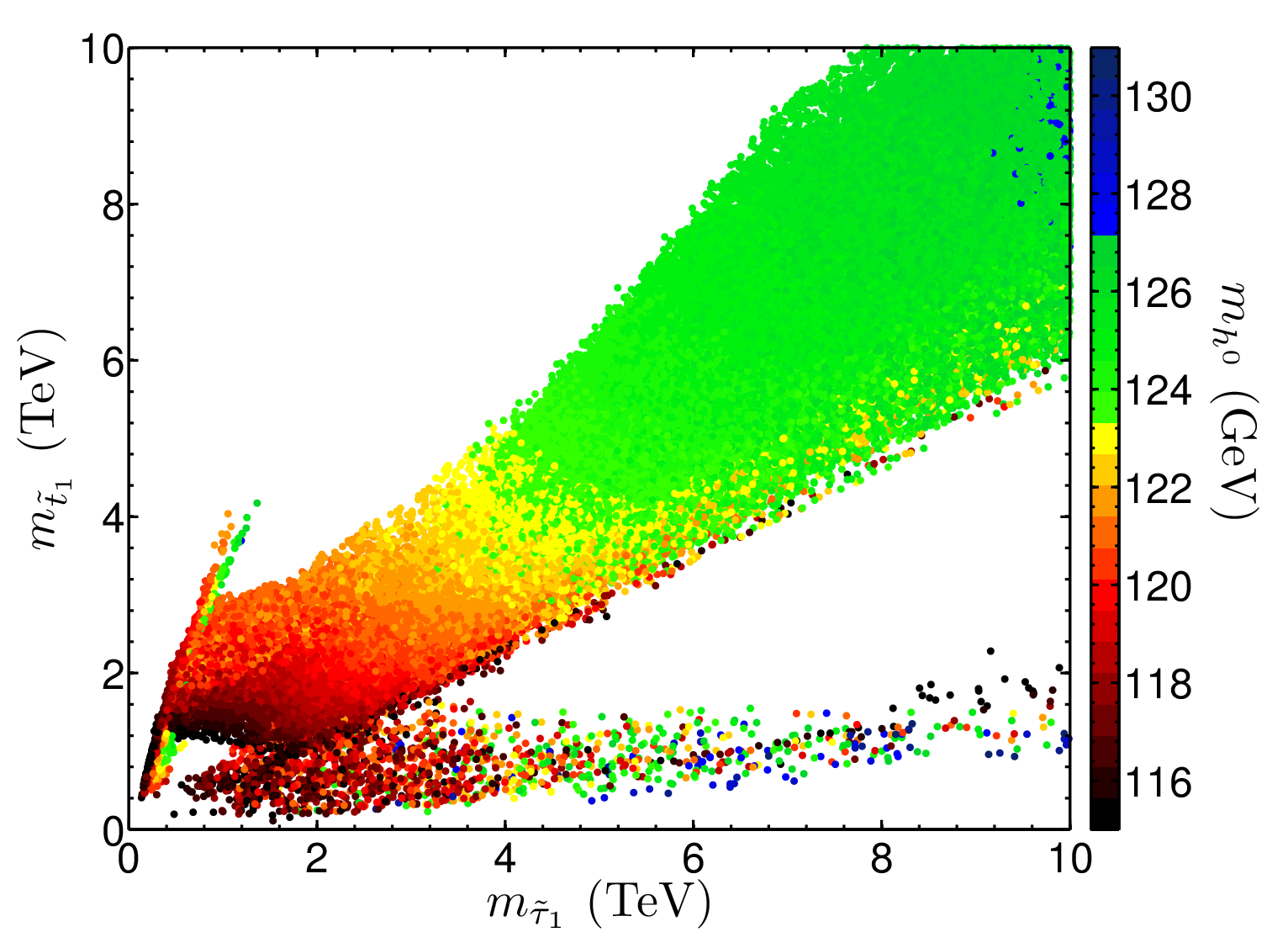}\hspace{.2cm}
\includegraphics[scale=0.1]{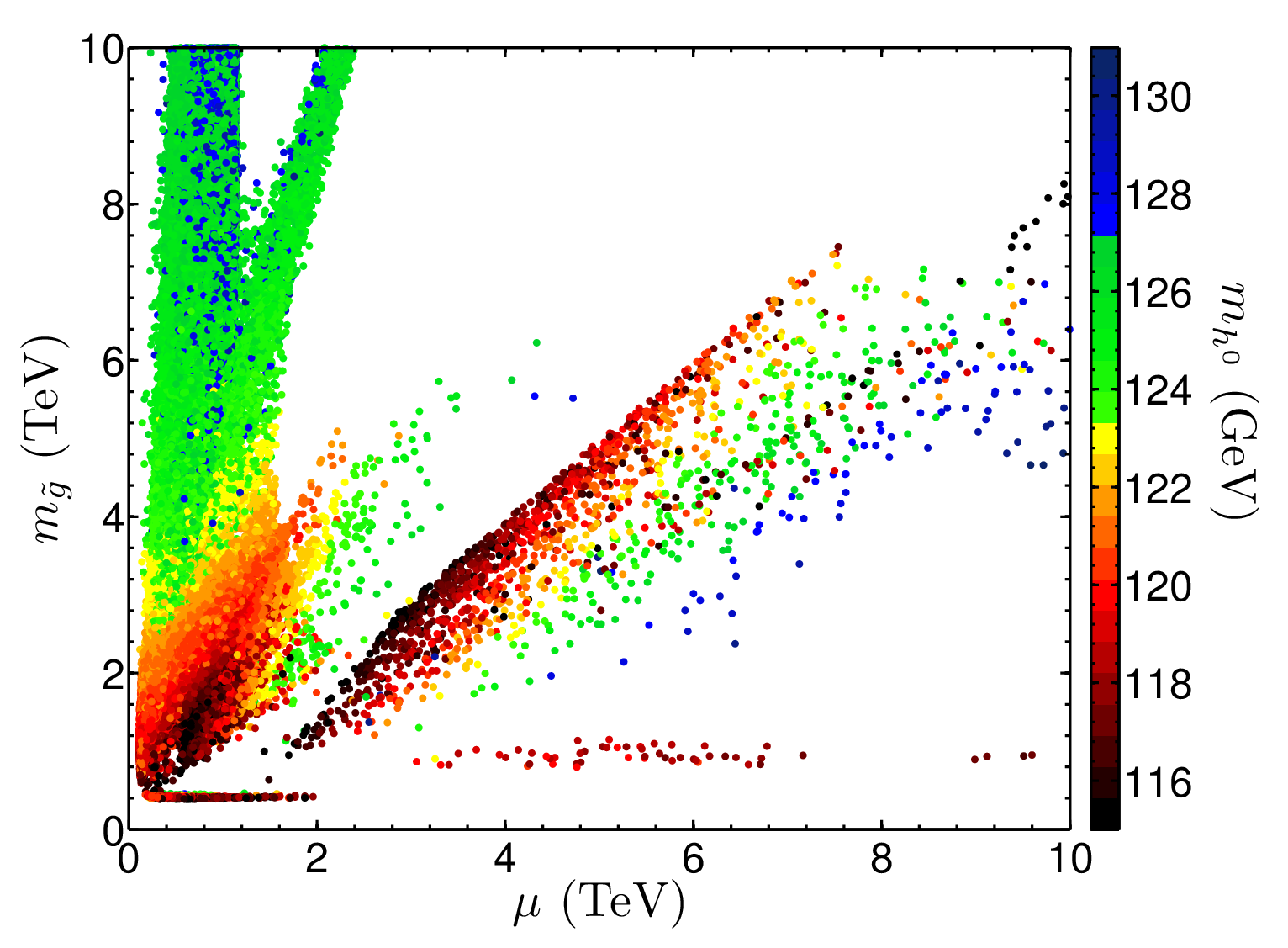}\hspace{.2cm}
\caption{(color online) 
Analysis is  based on the {\it general constraints} discussed in the text and for both scans of $m_0$.
Left panel: 
Exhibition of the stop vs the gluino mass in the mass window where both the stop and the gluino 
 masses run till $10\TeV$.
Middle panel: Exhibition of stop mass vs stau mass. Right panel: Exhibition of the 
gluino mass vs $\mu$. 
\label{sig}} 
\end{center}
\end{figure*}

\begin{table*}[t!]
\begin{center}
\begin{tabular}{|c|c|c|c|c|c|c|c|}
 \hline
& $\mh>115$ & $\mh>117$ & $\mh>119$  & $\mh>121$ & $\mh>123$ & $\mh>125$ & $\mh>127$ \tabularnewline\hline
\hline
$m_{H^0}\sim m_{A^0}$& 212 & 216 & 273 & 324 & 1272& 1517& 2730\tabularnewline
$m_{H^{\pm}}$ & 230 & 234 & 288 & 337 & 1275 & 1520 & 2732 \tabularnewline
\hline
$m_{\na}$ & 81 & 81 & 81 & 88 & 193 & 218 & 236 \tabularnewline
$m_{\cha}\sim m_{\nb}$  & 104 & 104 & 104 & 111 & 376 & 424 & 459 \tabularnewline
$m_{\g}$ & 800 & 800 & 803 & 803 & 1133 & 1264 & 1373 \tabularnewline
\hline
$m_{\ta}$ & 156 & 197 & 228 & 230 & 231 & 246 & 260 \tabularnewline
$m_{\sta}$ & 142 & 161 & 201 & 232 & 321 & 576 & 1364 \tabularnewline
$m_{\tilde{q}}$& 729 & 796 & 995 & 1126 & 1528 & 2235 & 2793 \tabularnewline
$m_{\tilde{\ell}}$  & 163 & 194 & 265 & 325 & 475 & 1631 & 2557 \tabularnewline
\hline
$\mu$   & 107 & 107 & 107 & 120 & 1418 & 1863 & 2293 \tabularnewline
\hline
\end{tabular} 
\begin{tabular}{|c|c|c|c|c|c|c|c|}
\hline
&$\mh>115$ & $\mh>117$ & $\mh>119$ & $\mh>121$ & $\mh>123$ & $\mh>125$ & $\mh>127$\tabularnewline
\hline
\hline
$m_{H^0}\sim m_{A^0}$ & 287 & 287 & 287 & 338 & 367 & 548 & 644 \tabularnewline
$m_{H^{\pm}}$ & 301 & 301 & 301 & 349 & 378 & 555 & 646 \tabularnewline
\hline
$m_{\na}$ & 91 & 91 & 91 & 91 & 91 & 91 & 256 \tabularnewline
$m_{\cha}\sim m_{\nb}$ & 104 & 104 & 104 & 104 & 104 & 104 & 261 \tabularnewline
$m_{\g}$ & 802 & 802 & 802 & 802 & 925 & 1006 & 1813 \tabularnewline
\hline
$m_{\ta}$ & 229 & 229 & 229 & 229 & 229 & 360 & 360 \tabularnewline
$m_{\sta}$ & 911 & 911 & 911 & 911 & 1186 & 1186 & 1186 \tabularnewline
$m_{\tilde{q}}$ & 4035 & 4035 & 4035 & 4035 & 4215 & 4493 & 4493 \tabularnewline
$m_{\tilde{\ell}}$ & 3998 & 3998 & 3998 & 4002 & 4085 & 4308 & 4308 \tabularnewline
\hline
$\mu$  & 118 & 118 & 118 & 118 & 138 & 140 & 251 \tabularnewline
\hline
\end{tabular}
\caption{Display of the lower limits on the sparticle masses as a function of a lower bound on the light Higgs mass for the mSUGRA models.  The top panel shows the sparticle lower bounds for the small $m_0$ scan and the bottom panel shows the sparticle lower bounds for the large $m_0$ sampling.  The {model} points in both cases pass the {\it general constraints} as well as an additional constraint that the gluino mass exceed $800\GeV$.
We note that the lower bound limits for the sparticles are not necessarily for the same model point. 
All masses are in GeV. A remarkable aspect of the analysis is that a stop mass as low as $300\GeV$ can be obtained for parameter points with $m_0>4\TeV$\label{spa_tab}. We further note that in this region 
one has the possibility of the first two neutralinos and the light chargino being degenerate 
as seen above when $\mu$ is smaller than the electroweak gaugino masses $\tilde m_1$ and $\tilde m_2$. }
\end{center}
\end{table*}


 There are some interesting correlations between the light Higgs and the sparticle spectrum.
 As noted already a larger light Higgs boson mass typically indicates a relatively heavier
 sparticle spectrum. 
 We give now a more quantitative discussion  using the two scans discussed in the previous section after imposing the {\it general constraints}.
 In Table~\ref{spa_tab} we present the lower limits on some of the sparticles as the light Higgs mass gets 
 progressively larger  between $\mh=115\GeV$ and $\mh=127\GeV$  {showing the results of  the two scans (upper and lower tables)}.  
 The top panel of the table is for the low value sampling of $m_0$, i.e. the scan with $m_0\leq4\TeV$, and the bottom panel is for the large value sampling of $m_0$, i.e. the scan with $m_0$ between $4\TeV$ and $100\TeV$.
Thus, after applying an additional $800\GeV$ gluino cut on the { models, for  the low $m_0$ scan we find that a } light Higgs boson mass of $\mh= 115\GeV$ allows for a lightest 
 neutralino mass of around $80\GeV$, but $\mh=125\GeV$ indicates a {lightest } neutralino mass of 
 around $220\GeV$.  The value of  220 \GeV ~is  consistent with independent constraints
coming from the search for squarks and gluinos at the LHC (see \cite{Akula:2011dd,Akula:2011ke}). 
For the  cases   $\mh=115\GeV$  and $\mh=125\GeV$ corresponding {masses} for the lightest chargino, $\cha$, ({degenerate with the} second lightest neutralino, $\nb$) are $100\GeV$ and $425\GeV$;
 for the gluino, $\tilde{g}$, $800\GeV$ and $1.3\TeV$; for the first and second generation squarks, $\tilde{q}$, $730\GeV$ and $2.2\TeV$,
 and for the first and second generation sleptons, $\tilde{\ell}$,  $150\GeV$ and $1.6\TeV$.  
Thus for 
  the low $m_0$ scan the shifts in lower limits are  dramatic for the gluino and for the
 first generation sfermions. The stop, $\ta$, and the stau, $\sta$, however,  continue to be relatively light. The $\sta$  mass, though is very sensitive to the higher mass bins in the light Higgs mass, i.e. bins greater then $123\GeV$.
 
For the large $m_0$ scan {the sparticle lower limits are modified in a significant way}. 
  {Most noticeably, the electroweak gaugino spectrum can remain light at higher Higgs mass 
  {relative to what one finds}
 in the more restrictive low $m_0$ scan.
 Further we observe that as the Higgs mass grows, the value of  $\mu$ can remain a few times the $Z$ mass, where as in the low $m_0$ scan this
does not occur.}
{In addition  we can see that the sfermion bounds do not change as drastically as the Higgs mass changes as they did  with the low $m_0$ scan,  and in particular the masses of the other Higgses $A^0, H^0, H^{\pm}$ can remain much lighter.}
 \begin{figure*}[htb]
\begin{center}
\includegraphics[scale=0.08]{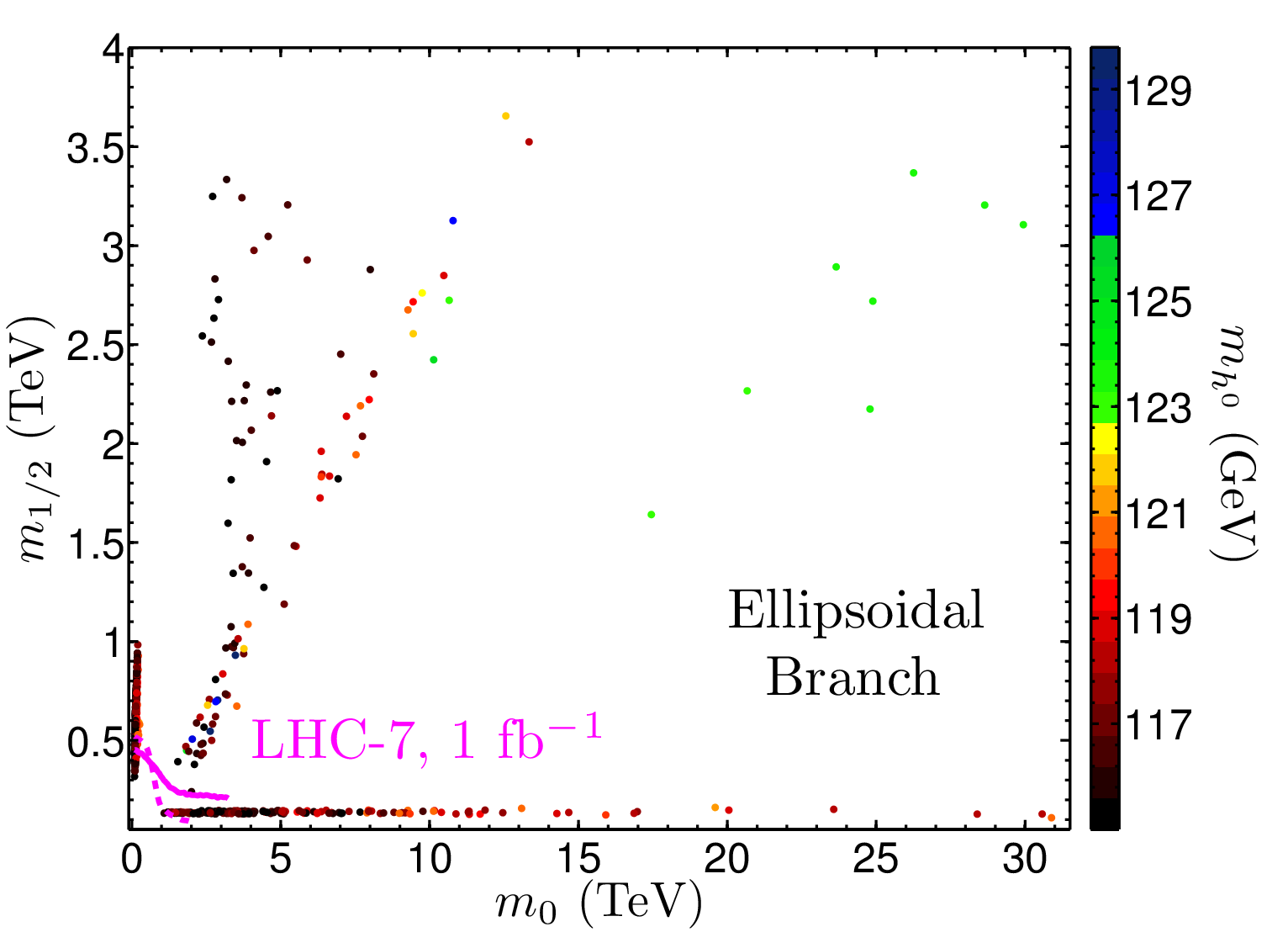}
\includegraphics[scale=0.08]{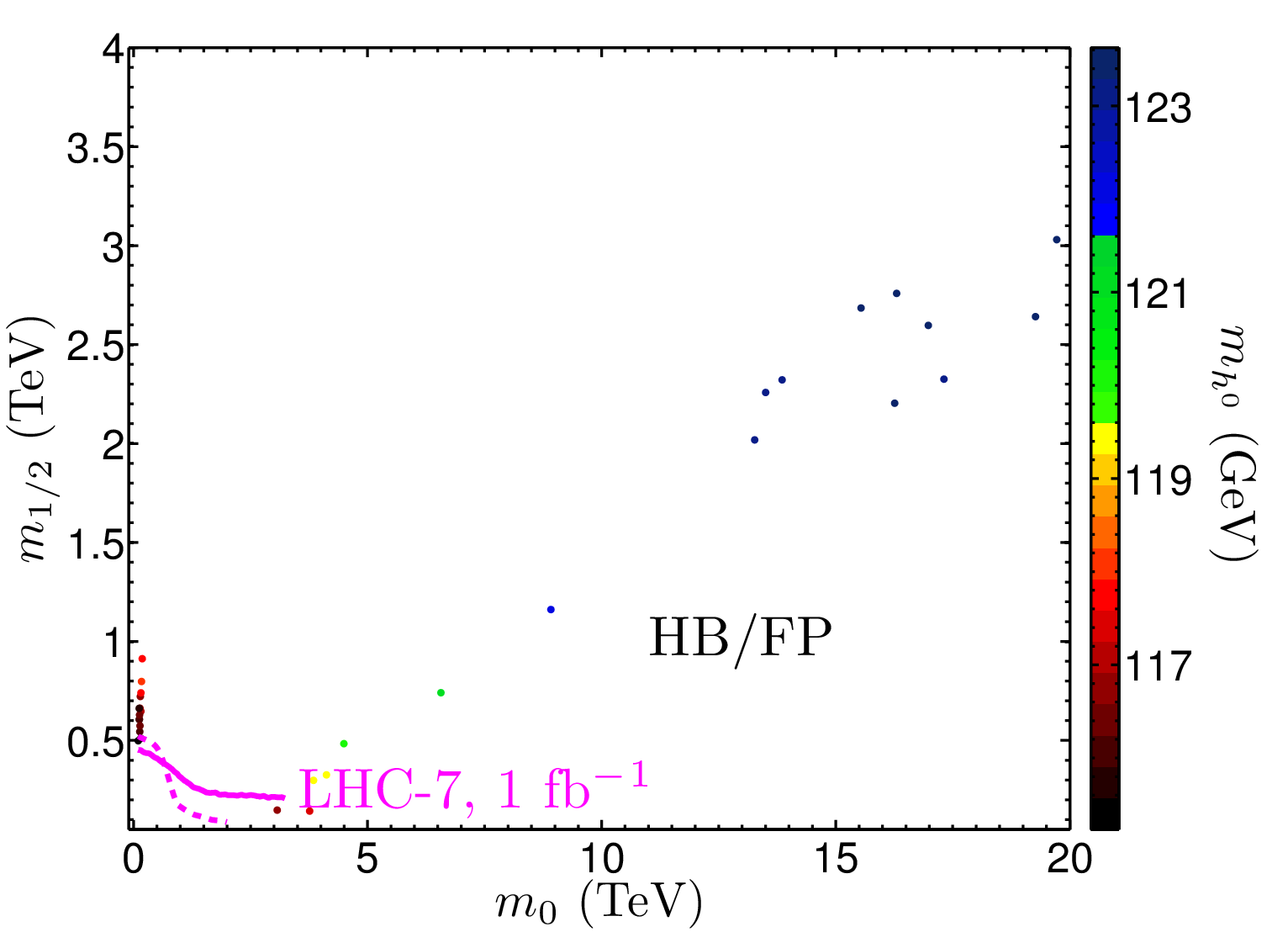}
\includegraphics[scale=0.08]{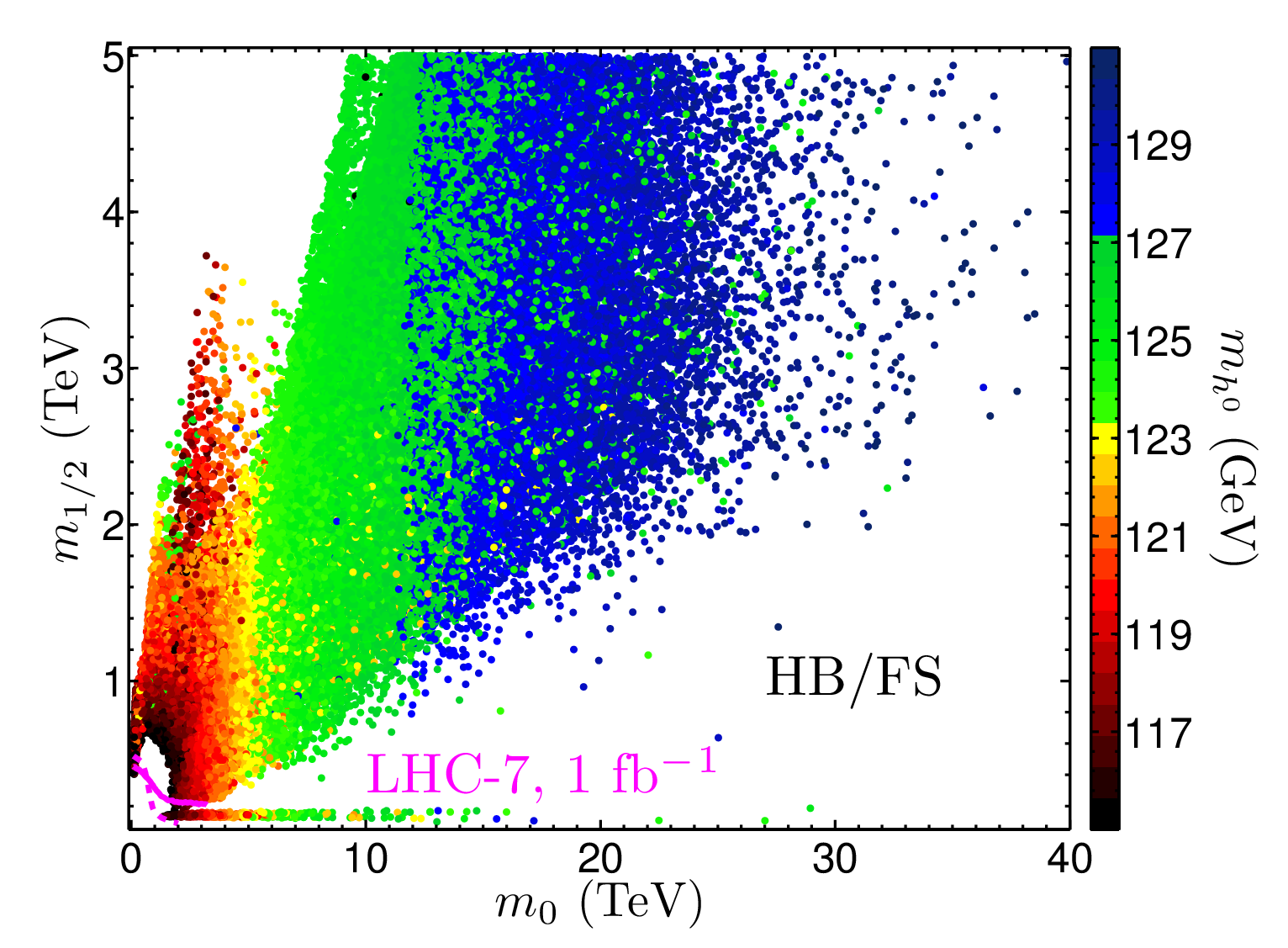}
\includegraphics[scale=0.08]{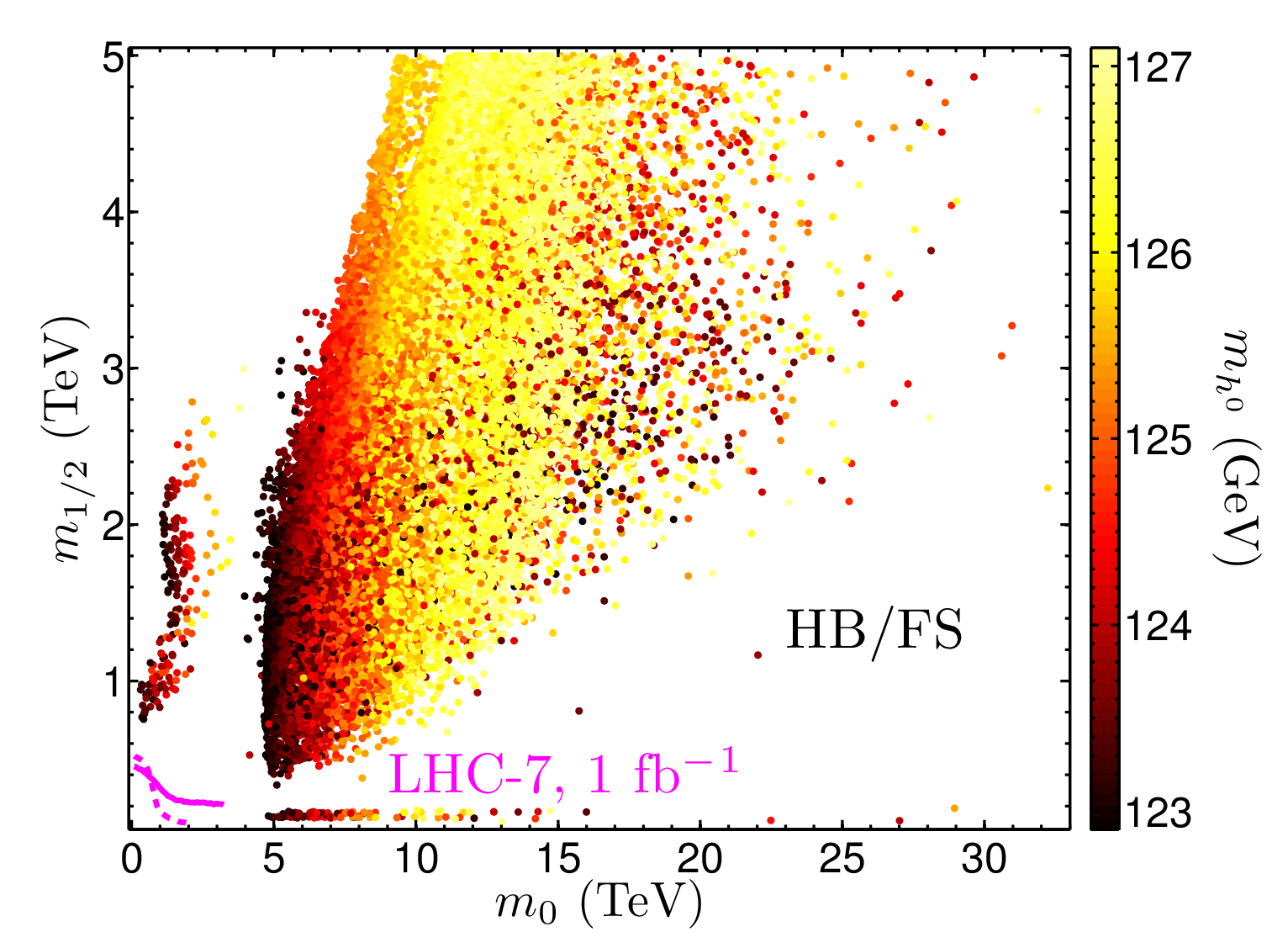}
\caption{
(color online) Analysis of the Higgs boson mass in Focal Regions. 
The analysis  is done for the {model} points that satisfy the low $m_0$ sampling and the {\it general constraints}.
Left: Shows the EB region with the light Higgs boson mass greater than $115\GeV$.  We see that the majority of these points are not in the heavy Higgs boson region.
Middle Left: Displays the HB/FP where we see that there are no Higgs masses greater then $120\GeV$.
In the right two panels we display the HB/FS (which include HB/FC) as follows:
in the middle right panel we exhibit the HB/FS {model} points for the Higgs mass range above $115\GeV$ and in the right panel we exhibit the HB/FS {model} points that have the light Higgs boson mass between $123\GeV$ and $127\GeV$.
\label{focal}}
\end{center}
\end{figure*}

  More graphically, in Fig.~\ref{sig} we compare ranges on the sparticle masses distributed by a light Higgs mass. 
Thus the left panel of Fig.~\ref{sig} gives a plot of the stop {mass} vs. the gluino mass and the middle panel gives a plot of the stop {mass  vs  the stau mass}.  
{These correlations of the light Higgs mass with the respective sparticle masses  show directly how a determination of the Higgs mass at the LHC will constrain the masses of
the R-parity odd particles.}
 The right panel of Fig.~\ref{sig} gives  a display of the gluino mass vs $\mu$ {(the Higgsino mass parameter at the scale $Q$ where electroweak symmetry breaking {occurs})}.
 Here one finds that a $\mu$, as small as a $200\GeV$, can generate a Higgs boson mass
 up to about $122\GeV$. {However, the larger Higgs masses, i.e., Higgs masses above $125\GeV$
can also have $\mu$ of size that is sub-TeV}.  Thus, one can have a heavier Higgs, 
{scalars in the several TeV region},  but still have a 
 light $\mu$ {\cite{Chan:1997bi,fln-higgs,Feldman:2011ud}}.  
\section{Hyperbolic Branch of  REWSB \newline and Focal Surfaces}
  It is known that the radiative electroweak symmetry breaking carries in it a significant 
 amount of information regarding the parameter space of SUGRA models.
  Thus  REWSB allows for a determination of  $\mu^2$ in terms of the 
  soft parameters~\cite{Nath:1997qm,Chan:1997bi} (for further works see~\cite{bbbkt})
  so that {the breaking of electroweak symmetry is encoded in the following expression }
 \beqn
 \mu^2  = &-&\frac{1}{2}M_Z^2 +  m^2_0  C_1+ A^2_0 C_2\nonumber\\
 & +&
 m^2_{1/2} C_3+ m_{1/2}
A_0 C_4+ \Delta \mu^2_{\rm loop}~,
\label{2}
\eeqn 
 where $C_i$, $i$ running from $1$ to $4$,  {depend}  on the top mass, $\tan\beta$ and $Q$.
 It was shown in~\cite{Chan:1997bi} 
  that one can classify regions of  Eq.~(\ref{2}) in the following two 
 broad  classes: the Ellipsoidal Branch, denoted EB, where $C_1>0$, and the Hyperbolic Branch, denoted HB,  where $C_1\leq 0$.
 More recently in~\cite{Akula:2011jx} it was shown that  HB  can be further classified into three regions.  One such region was defined as the Focal Point, HB/FP, where $C_1=0$.  It was further shown that 
the HB/FP limits to the Focus Point~\cite{Feng:1999mn} when $\tan\beta\gg1$.  Another region defined was the 
 Focal Curve, HB/FC, where
 $C_1<0$ and two soft parameters are free to get large, i.e., either $m_0, A_0$  or $m_0, m_{1/2}$.  The last region was defined to be the Focal Surface, HB/FS, where $C_1<0$ and three soft parameters were free to get large, i.e.,
 $m_0, A_0, m_{1/2}$. It was further shown in~\cite{Akula:2011jx} that HB/FC was a subset of HB/FS and that the HB/FP was mostly
 depleted after imposing constraints from flavor physics, WMAP, sparticle mass lower limits
and LHC-7. However, other regions 
 of the parameter space were found to be well populated.

 In Fig.~\ref{focal} we give an analysis of the Higgs mass ranges lying on the EB and on the Focal Regions.  In the top two panels we consider the Higgs mass range upwards of $115\GeV$.
 The  left panel is  for the Ellipsoidal Branch and the middle left
 panel is  for the Focal Point region.
    In the EB region one finds that the majority of light Higgs boson masses do not exceed $124\GeV$, while
 in the HB/FP region the Higgs masses do not get beyond $120\GeV$ except perhaps 
 for some isolated points.   Further the HB/FP region is highly depleted
 {as can be  seen by the paucity of allowed model points in the middle left panel of  Fig.~\ref{focal}}.
 The largest Higgs boson masses are achieved on HB/FS, which includes HB/FC, shown in the right two panels of 
 Fig.~\ref{focal} where  the region above  a Higgs boson mass of $115\GeV$ (middle right) and 
 between $123\GeV$ and $127\GeV$ (right) are shown. The right panel shows that the Higgs mass region 
 within a few  $125\GeV$ is well populated.


\begin{figure*}[t!]
\begin{center}
\includegraphics[scale=0.14]{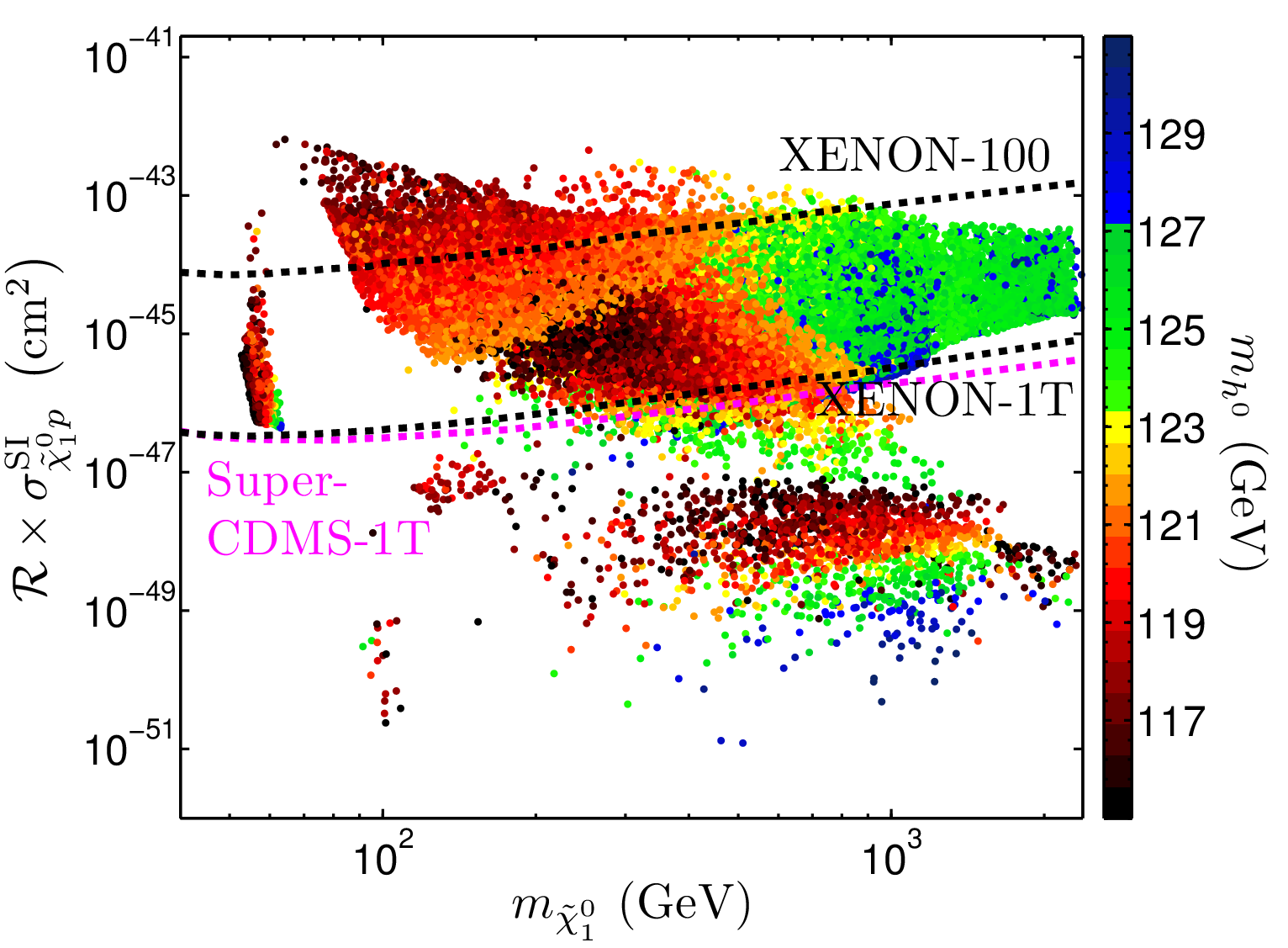}
\includegraphics[scale=0.14]{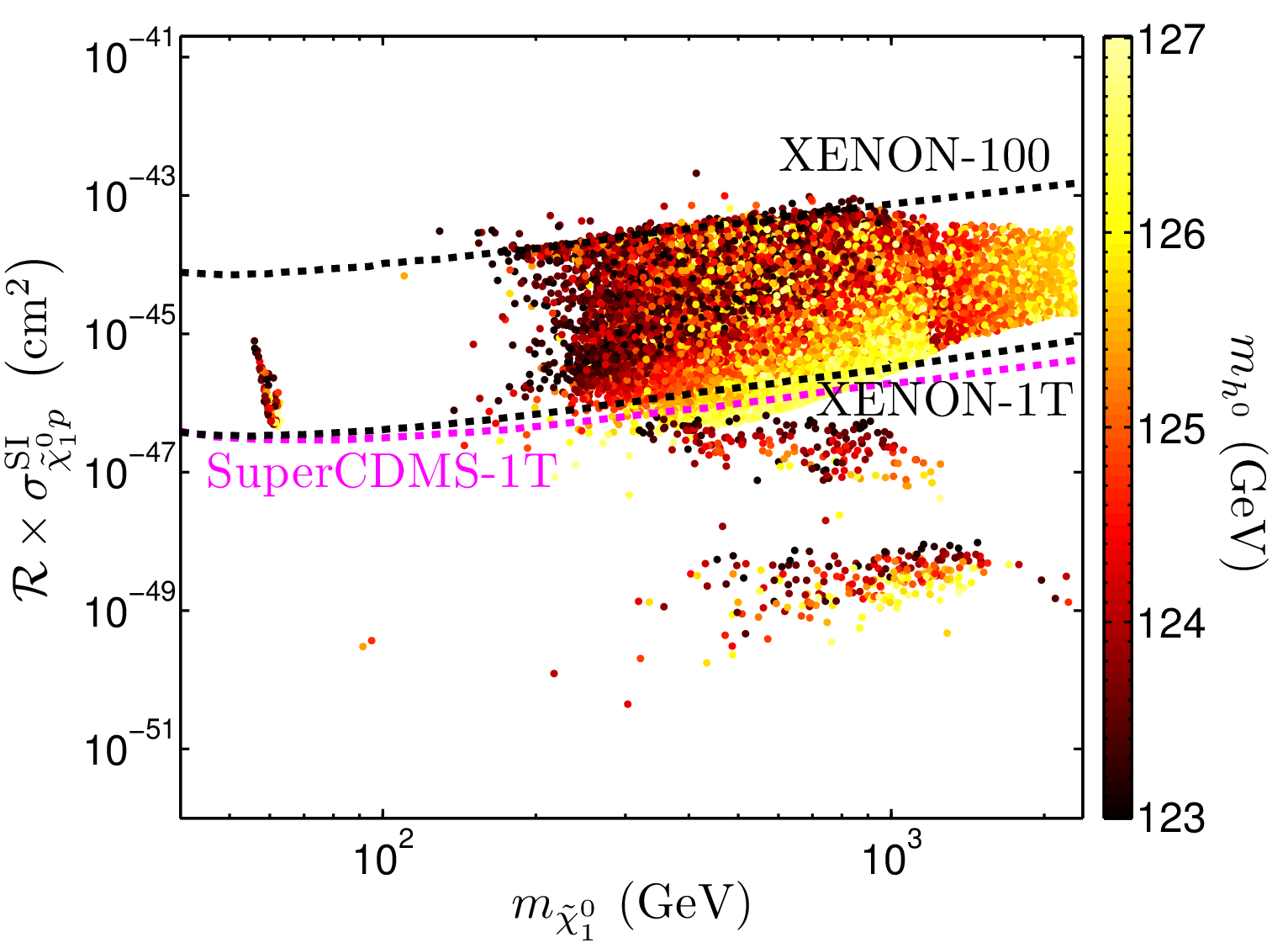}
\caption{(color online) 
 Exhibition of proton-neutralino spin-independent cross section against the neutralino mass.  {Here we see that  models with a Higgs Boson mass in the range consistent with  
 the results from LHC-7} will be probed in the next round of dark matter experiments.
In  the plots the proton-neutralino spin-independent cross section was corrected by $\mathcal{R}\equiv\left(\Omega h^2\right)/\left(\Omega h^2\right)_{\rm WMAP}$ to allow for multicomponent dark matter.
The analysis is done for the {model} points passing the {\it general constraints} from the low $m_0$ sampling. The left panel gives the full light Higgs boson mass range, i.e. $115\GeV$ to $131\GeV$ and the right panel only deals with the sensitive region between $123\GeV$ to $127\GeV$.
\label{dark}}
\end{center}
\end{figure*}

 \section{Higgs boson  and dark matter}
 There is  a strong correlation between the light Higgs mass  and dark matter. 
 It has already been pointed out that annihilation via the Higgs pole can generate
 the relic density to be consistent with WMAP (see the first paper of \cite{LHC-7}). In this case the neutralino mass
 would be roughly half the light Higgs boson mass. For heavier neutralino masses
 other annihilation mechanisms become available. We would be interested in the cases which include
 large $m_0$ and specifically in the spin independent proton-neutralino cross section in this domain.
 For this case when $m_0$ is large
 the $s$-channel squark exchange  which contributes to the spin independent proton-neutralino 
 cross section becomes suppressed while
the   $t$-channel Higgs exchange  dominates.  The scattering cross section in this case is given by
\beqn
\sigma^{\rm SI}_{\na N}  =  \left(4 \mu^2_{\na N}/\pi\right) \left(Z f_p +(A-Z)f_n\right)^2~.
\eeqn
Here 
$f_{p/n}$ $ =$ $\sum_{q=u,d,s} f^{(p/n)}_{T_q} {C}_q \frac{m_{p/n}}{m_q} $ $+$ $ \frac{2}{27} f^{(p/n)}_{TG} $ $\sum_{q=c,b,t} {C}_q \frac{m_{p/n}}{m_q}$, 
where  the form factors $f^{(p/n)}_{T_q}$  and  $f^{(p/n)}_{TG} $  are given in \cite{Ibrahim,Ellis,belanger} and  the couplings $C_i$ are given by \cite{Ibrahim,Ellis}
\begin{eqnarray}
{C}_q   = &   -& \frac{g_2 m_{q}}{4 m_{W} \delta_3} \left[  \left( 
g_2 n_{12} - g_Y n_{11}] \right) \delta_{1} \delta_4\delta_5 \left( - \frac{1}{m^{2}_{H}} + 
\frac{1}{m^{2}_{h}} \right) \right. \nonumber \\
& +&  \left. \left(g_2 n_{12} - g_Y n_{11} \right)  \delta_{2}\left(  \frac{\delta_4^{2}}{m^{2}_{H}} +\frac{\delta_5^{2}}{m^{2}_{h}}\right) \right]~.
\end{eqnarray} 
 For up quarks one has 
$\delta_i = (n_{13},n_{14}, s_{\beta},s_{\alpha},c_{\alpha})$ and  for down quarks 
$\delta_i=
(n_{14},-n_{13}, c_{\beta},c_{\alpha},-s_{\alpha})$,  where $i$ runs from $1$ to $5$, $\alpha$ is the neutral Higgs mixing 
parameter, $n_{1j}$ is the neutralino eigencontent, $c_{\alpha}$ denotes $\cos\alpha$ and $s_{\alpha}$ denotes $\sin\alpha$. 
The  above approximation holds  over  a significant part of the parameter space specifically
for large $m_0$ and we have checked that it compares well with the full analysis where the 
  full theory calculation is done with {\sc micrOMEGAs}. In the analysis  work presented here, however,
  we exhibit only the results of the full analysis. 
 In  Fig.~\ref{dark}  we give a plot of the 
proton-neutralino spin-independent cross section, $\sigma_{\na p}^{\rm SI}$ times $\mathcal{R}$
plotted as a function of the neutralino mass where we have corrected $\sigma_{\na p}^{\rm SI}$   
    by a factor $\mathcal{R}\equiv\left(\Omega h^2\right)/\left(\Omega h^2\right)_{\rm WMAP}$ to take into account the possibility of multicomponent dark matter.  
The points are  shaded according to the Higgs boson masses and we show the XENON-100~\cite{xenon} exclusion curve as well as the  XENON-1T~\cite{futureXENON} and the
 SuperCDMS~\cite{futureSCDMS} projections.

It is important to observe that when the Higgs mass region
 $123\GeV$ to $127\GeV$ is considered, nearly all of the mSUGRA parameter 
points that lie in this region which are 
also consistent with the {\it general constraints} (from our  low $m_0$ and high $m_0$ scans) 
give rise to neutralino mass and proton-neutralino spin-independent cross section 
(scaled by $\mathcal{R}$), that lies  just beyond what the most recent results from 
the XENON collaboration have probed. However, a vast majority of this region is 
projected to be explored by XENON-1T and SuperCDMS. This point is clearly 
 seen in the right panel of Fig.~\ref{dark}.

   
 \vspace{-.5cm}
\section{Conclusion}
Recent data from  LHC-7 indicates  a narrow window on the light Higgs mass.
This  allowed mass window 
 is consistent with the range predicted by SUGRA models and specifically by 
 the mSUGRA model.  Here we discussed 
the implications of the indicated mass range for the light Higgs mass for the sparticle mass spectrum 
and
for dark matter.
Using the allowed Higgs mass range above $115\GeV$    the corresponding ranges for the soft masses and couplings, as well
as the ratio of the vacuum expectation values of the Higgs doublets and the Higgsino mass parameter
 were found.  We then investigated the ranges for the sparticle masses correlated to the predicted value
of the Higgs Boson mass, specifically
for the chargino, the neutralino, the gluino, the stop, the stau, for the first and second generation squarks and
sleptons {and for the heavier Higgs of  the minimal supersymmetric standard model, i.e., 
the CP odd Higgs $A^0$, the CP even Higgs $H^0$, and the charged Higgs $H^{\pm}$.}

{Our conclusions are}   that the largest Higgs masses are realized on the Focal Surface
{of the Hyperbolic Branch} 
of radiative electroweak symmetry breaking.   {We also point out that low values
of $\mu \sim 150 \GeV$ are consistent with heavy squarks and sleptons in the 10 TeV region or  larger.   We find that 
 $\mh~\in (123-127)\GeV$   does allow for light third generation stop as low as  $m_{\tilde t_1} > 230\GeV$, though
  the second generation squarks are at least $m_{\tilde q} > 1.5\TeV$  and second generation sleptons are at least 475 GeV.}
  {
    Thus,  the restriction of the light Higgs boson to the mass window  $\mh~\in (123-127)\GeV$ 
 provides further  constraints on the sparticle spectrum that are complimentary to the direct
 searches for sparticles at the LHC.  }

{ Further, 
we find precise predictions for dark matter  if 
the light Higgs boson mass lies between $123\GeV$ and $127\GeV$.}
For these light Higgs boson masses, the corresponding range 
of the lightest neutralino  mass  would be 
accessible in the next generation of direct detection dark matter experiments.
The light Higgs boson in the $123\GeV$ and $127\GeV$ range was shown to be generic
for the case of heavy scalars in minimal supergravity with {$|A_0/m_0|\sim\mathcal{O}(1)$.}

\begin{acknowledgements}
This research is  supported in part by grants  NSF grants PHY-0757959 and PHY-0969739,  
 by the DOE grant  DE-FG02-95ER40899, by the Michigan Center for Theoretical Physics, 
and by  TeraGrid  grant TG-PHY110015.
\end{acknowledgements}

\noindent Note Added: After this work was finished  the papers of \cite{newpapers} appeared which also investigate
the implications of the recent Higgs limits for the mSUGRA model.


\end{document}